\documentclass[aps,pra,preprint,groupedaddress,showpacs,showkeys]{revtex4}

\usepackage{graphicx}
\usepackage{doublespace}

\usepackage{amssymb}
\usepackage{amsmath}
\usepackage{amsthm}

\def\prd{Phys. Rev. D}

\def\w{\omega}

\def\th{t_0}
\def\*{\cdot}

\def\d{\partial}
\def\de{\delta}

\def\ba#1{\overline{#1}}
\def\v#1{\mathbf{#1}}
\def\uv#1{\underline{\mathbf{#1}}}
\def\t{\tau}
\def\ra{\rightarrow}

\def\ax{\uv{x}}
\def\ay{\uv{y}}

\def\PD{\text{PD}}

\def\wq{\w_\v q}
\def\X{\ax}
\def\s{ \sigma}

\begin{document}


\title{Balanced homodyne detectors in QFT}


\author{Piotr Marecki}
\email[]{pmarecki@gmail.com}
\affiliation{ITP, Universitaet Leipzig, Postfach 100 920, D-04009 Leipzig, Germany}


\date{\today}

\begin{abstract}

By examining BHDs from the general quantum-field-theoretical point of view we derive a number of generalizations of the standard analysis of their response. In particular, by allowing for interactions restricted in time (by a smooth function) we present general expressions for BHD's output (in which the usual, simplifying limits can but need not to be taken). Moreover we point out the need for non strictly-monochromatic local oscillators (i.e. the need for ``pulsed'' ones) in order to have well-defined QFT-observables (the products of which, eg. have finite vacuum expectation values). Furthermore we show how the analysis of the detectors generalize to situations with BHDs in waveguides, Casimir cavities, or other time-independent but inhomogeneous $\epsilon$ and $\mu$. The general treatment also allows us to comment on important QFT features of the detector-observables such as locality (i.e. commuting for causally separated measurements) and non-null vacuum response. This leads to the conclusion that balanced homodyne-type detectors (and not single-photodiode-like ones) are the appropriate tools in testing intriguing QFT predictions like negative (sub-vacuum) energy-densities. Finally by recalling results on field-autocorrelation functions for QFT in Casimir cavities we show that interesting effects (large reductions of fluctuations) are to be expected if a version of BHDs were to be placed in such cavities.
\end{abstract}


\maketitle

\section{Introduction}
In quantum field theory the vacuum is not as empty as in the classical physics. The vacuum expectation values of positive operators, for instance of the square of the smeared electric field, are non-vanishing and there are plenty of states with the expectation values lower then the vacuum ones. These states are locally ``darker than vacuum''. While the understanding of this phenomenon progressed constantly over the last decades, with a number of surprising results having been discovered\footnote{See, for example, \cite{FR} and the references therein.}, attempts were also made to see this non-classical behavior in experiments. Naturally, in order to see sub-vacuum fluctuations it is at least necessary to have detectors capable of seeing and quantifying the vacuum fluctuations. Such detectors, known as balanced homodyne detectors (BHD) with local oscillators, were first proposed in the context of quantum optics by Chen and Yuan \cite{first}. With their help the so-called squeezed states of light are seen to exhibit regions with sub-vacuum electric field fluctuations (eg. \cite{pulsed}).

  In this paper we reconstruct the quantum-field-theoretical observables corresponding to photodiodes and balanced homodyne detectors with local oscillators and discuss their quantum-field-theoretical properties. The paper is organized as follows: in the part A of the second section we investigate the interaction of a simple quantum system with the quantum electromagnetic field and derive a formula for the response of a photodiode. The part B deals with an arrangement of two photodiodes known in quantum optics as the balanced homodyne detector. This part contains four subsections dealing with generalizations of the standard derivation of the observable corresponding to measurements on BHDs. In particular we show in which asymptotic sense the expectation values of the output of BHDs, and its variance, are related to the one- and two-point functions of the quantum electromagnetic field (subsections 1 and 4). We generalize these results to the case of non-monochromatic local oscillators (subsection 2) and to the case of BHDs placed in inhomogeneous environments (waveguides, Casimir cavities, inhomogeneous media; section 4). The utility of these generalizations is shown in the appendix A, where we analyze the expected response of BHDs placed in Casimir cavities (consisting of two parallel, perfectly conducting plates), and argue that substantial, time-independent, reductions of fluctuations (aka sub-vacuum expectation values of the square of the electric field operator) are expected. The third section contains a number of comments on the nature of  BHD-measurements and provides a relation to more abstract QFT results (vacuum response, causality and the Reeh-Schlieder theorem).

\section{Detectors}

\subsection{Photodiode}
Let us start the discussion with a photodiode, on which a photodetection process takes place in a semiconductor region (usually the I region of a PIN diode). If light is shed upon this region, electron-hole pairs
are created. Upon creation, these charges propagate towards and eventually reach the opposite sides of the
diode, where they are collected. In this paper we shall model the detection as an excitation of an electron, initially in a relatively well-localized bound state $\psi_0$ (eg. of an isolated donor atom), to the continuum of excited states, $\psi(\v q)$.

We will adapt here the of standard treatment of the detection process, presented eg. in the complement $A_{II}$ of \cite{CCT}, or in the
section 4.2 of \cite{SZ}. Let the dynamics of the electron be generated by the Hamiltonian
\begin{equation}
H=H_0+V_{int}
\end{equation}
with
\begin{equation}
H_0= p^2/2m^\star +V(\v x)
\end{equation}
$\psi_0$ and $\psi(\v q)$ being eigenstates of $H_0$, and the dipole-approximation interaction-Hamiltonian
\begin{equation} \label{f_dipol}
V_{int}=e x^i \,  E_i(t,\v x),
\end{equation}
where $e x^i=d^i$ is the dipole-moment operator and $E_i(t,\v x)$ is the (quantum) electric-field operator. We shall assume, that this field-operator  can be replaced by the electric-field operator taken \emph{at the point}, $\ax$, where the donor-atom resides,
\begin{equation}\label{hamiltonian}
 E_i(t,\underline{\v x})=\int d^3x\, E_i(t,\v x) \de(\v x-\ax).
\end{equation}
This simplifying assumption will be argued unnecessary for the BHD in section \ref{rem2}. The evolution will be determined from the first order time-dependent perturbation theory. We shall use the formulation, \cite{pm,bf}, in which the interaction Hamiltonian is multiplied with a smooth function of time,  $g(t)$,  equal to one in the time
interval during which the measurement takes place\footnote{An
approach of this type has also been adapted by Ottewill
and Davies, \cite{DO}.}, and vanishes rapidly just outside of this interval.
The probability of exciting the electron by a given
initial state $S$ of the quantum field is just the expectation value of the (time-evolved) projection operator on the excited states of the electron, say $P_e\otimes \mathbf 1$, taken with respect to the state $\psi_0\otimes S$ of the full system. The well-known first order perturbative expression for this probability is
\begin{multline}
  \PD(g,\ax)=\langle \psi_0\otimes S|U^*_g\, (P_e\otimes \mathbf 1) \, U_g |\psi_0\otimes S\rangle=\\ \int_{-\infty}^{\infty} g(\tau)g(s)\ ds\ d\tau
\int d\mu(\v q)\,  \bigl(\psi_0, e^{iH_A s}
  \ d^i\ e^{-iH_As} \psi(\v q)\bigr)\bigl(\psi(\v
  q),e^{iH_A\tau}
  \ d^j\  e^{-iH_A \tau}\psi_0 \bigr) \ \left\langle E_i(\tau,\uv x)E_j(s,\uv
  x)\right\rangle_S,
\end{multline}
where $U_g$ is the evolution operator with the interaction controlled by $g(t)$, and the measure $d\mu(\v q)$, specifies the final, scattering states of the electron. The above expression can be written in a compact form:
\begin{equation}
  \PD(g,\ax)= \int d^2g\,
 d\mu(\v q)\,  e^{i\w_\v{q}(\tau-s)}\, \overline{d^i(\v q)}d^j(\v q) \, \left\langle E_i(\tau,\uv x)E_j(s,\uv
  x)\right\rangle_S,
\end{equation}
where symbolically $d^2g\equiv g(\tau)g(s)\, ds\, d\tau $, while $\w_\v{q}=E(\v q)-E_0$ stands for the unperturbed (w.r.t. $H_0$) energy difference between the excited and the ground state of the electron and $d^j(\v q)=\bigl(\psi(\v q), \, e\, x^j\, \psi_0 \bigr)$. It is convenient to introduce a symbolic star notation, such that
\begin{equation}
   \PD(g,\ax)=\int_{-\infty}^{\infty} d^2g\,
    \left\langle E(\tau,\uv x)\star E(s,\uv
  x)\right\rangle_S
\end{equation}

\subsection{Balanced homodyne detector (BHD) with a local oscillator}

The primary purpose of a BHD with a local oscillator (LO) is to investigate the states of the radiation field, which alone hardly trigger photodiode's response, eg. vacuum/ground states. Consider the
setup presented on the figure 1. The observable
corresponding to the charge collected at $V$ is the difference of
the outputs of the photodiodes:
\begin{equation}
J=\PD(g,\ax)-\PD(g,\ay).
\end{equation}

\begin{figure}[ht]\centering
  \includegraphics{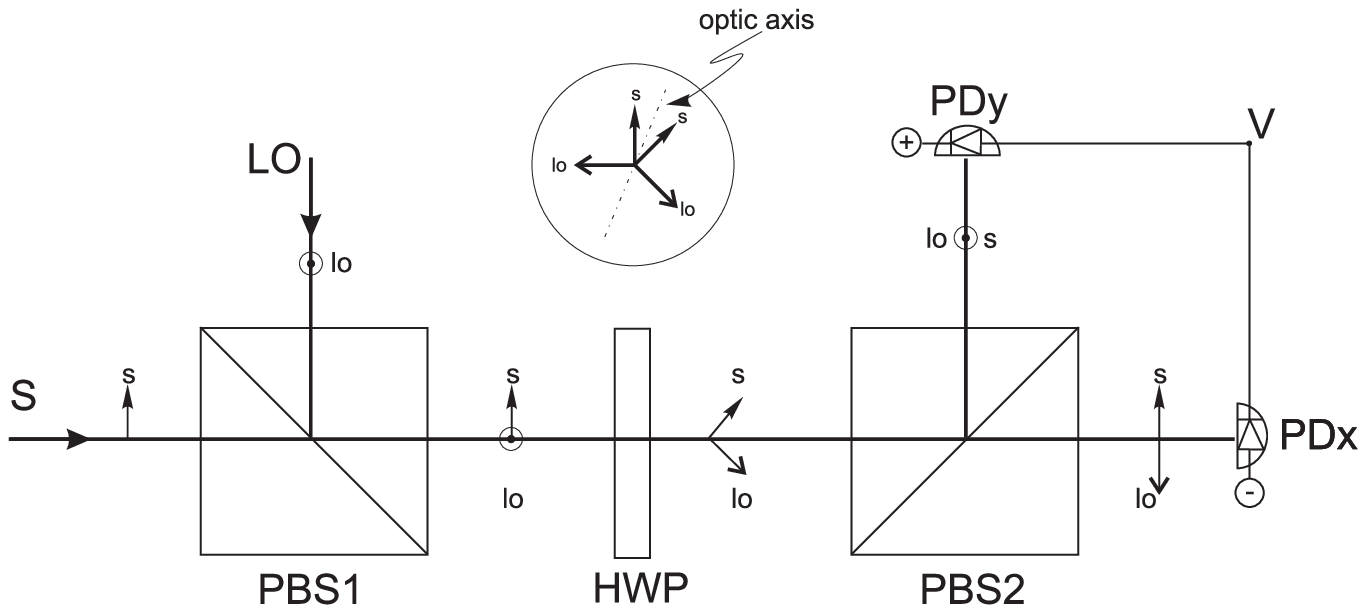}
  \caption{Balanced homodyne detector with a local oscillator. The linearly polarized signal
  field, $S$, (if present) is ``blended'' with a coherent state (LO),
   which is polarized orthogonally to $S$, on the polarizing
  beam splitter (PBS1). The half wave plate (HWP), reflects the
  planes of polarization with respect to its optical axis, thereby
  inducing $\pi/4$ shift of the plane of polarization of the
  signal field $S$. The subsequent PBS2 separates the two orthogonally polarized signals,
  which are detected at the photodiodes $PDx$ and $PDy$. The charge collected at $V$ provides a measure of the
  expectation value of the observable $J$ (and of its higher moments). Note, that the setup is
  arranged in such a way, that if $S$ happened to be a monochromatic coherent
  state, it would be phase-matched to the LO at the point $x$, but
  shifted in phase by $\pi$ at the point $y$.}
  \label{bhd}
\end{figure}

In experiments with BHD  the state
under investigation is ``blended'' with a strong, precisely controlled coherent
state (LO). Let $S$ denote the state of the radiation field under investigation (that is: the state of the quantum field \emph{without} the local oscillator present). In a typical experimental setup (fig. 1) the polarising beam splitter, PBS1, superposes
physically the state $S$ and the LO.  The blending has a ``coherent
character'' and therefore we adapt one of the definitions of the
coherent state to describe it:
\begin{equation*}
  \langle P[E_i(t,\v x)] \rangle_{(S,F)}=\langle P[F_i(t,\v x)+E_i(t,\v
  x)]\rangle_S,
\end{equation*}
where
\begin{equation}
F_i(t,\v x) \stackrel{def}{=} \langle E_i(t,\v
  x)\rangle_F
\end{equation}
denotes the electric field of the LO. The symbol $P$
above stands for an arbitrary observable (polynomial) constructed out of the
(smeared) electric field operators, while
 $(S,F)$ denotes the state resulting from the
blending of the local oscillator coherent field into the state $S$
under consideration. We choose the LO field to be monochromatic and linearly polarized, so that at every point $\ax$ we may write
\begin{equation}
F_i(\tau,\ax )=K_i \left[e^{-i\w ( \tau-\th)}+e^{i\w( \tau-\th)} \right],
\end{equation}
with a real vector $K_i$ and the frequency $\w>0$. The phase $\th$, which depends on the point $\ax$, can be varied at will in experiments. Per definition the detector is arranged (balanced) is such a way that\footnote{Up to a necessary rotation, see figure 1 and the note below.}
\begin{equation}
F_i(\tau,\ax )=-F_i(\tau,\ay ).
\end{equation}
The expectation value of $J$ for the detector balanced in this way is
\begin{multline}\label{av_J}
  \langle J\rangle_{(S,F)}=\langle J\rangle_S+ \int d^2g\,
  \left\{[F(\tau, \ax)\star \langle E(s,\ax)\rangle_S+ \langle E(\tau,\ax)\rangle_S \star F(s, \ax)] -\right.\\
  -\left.[F(\tau, \ay)\star \langle E(s,\ay)\rangle_S+\langle E(\tau,\ay)\rangle_S \star F(s, \ay) ] \right\}
\end{multline}
(the terms proportional to $F^2$ cancel.) The quantum electric field is
\begin{equation}
E_i(s,\ax)=-\d_s\, A_i(s,\ax)=-\d_s\,
\frac{1}{\sqrt{2\pi}^3}\int\frac{d^3p}{\sqrt{2p}}\left[ e^{-ips}
 a_i(\v p)+e^{ips}a^*_i(\v p)\right],
\end{equation}
where $p=|\v p|$ and
\begin{equation}
[a_i(\v p),a^*_j(\v k)]=\de(\v p -\v k)\, (\de_{ij}-p_ip_j/|\vec p|^2).
\end{equation}
It can also be written briefly (by defining appropriate $b_i(\v p)$):
\begin{equation}\label{electric_field}
E_i(s,\ax)=\int d^3p\, \left[ e^{-ips}
 b_i(\v p)+e^{ips}b^*_i(\v p)\right].
\end{equation}

Let us now symbolically perform the $\t$ and $s$ integrations in the operators of the form $\int d^2 g\, F\star E$, the expectation values of which appear in eq. \eqref{av_J}. The integration leads to terms involving
\begin{equation}
 \hat g(\w_\v q\mp \w)\, \hat g\left(\mp
p-\w_\v q\right).
\end{equation}
Of these, clearly the term $\hat g(\wq-\w)\hat g(p-\wq)$ dominates the other three, because $\w,\wq,p$ are all greater than zero, and the Fourier transform of $g(t)$
is concentrated around the zero frequency with a rapid decay away from it. This, as explained in the appendix A,
essentially enforces the restrictions $\w_\v q=\w$ and $p=\w$  in the integrations over
$d\mu(\v q)$ and $d^3p$ respectively.

\subsubsection{Asymptotic response of BHDs}
At this point let us make a simplifying assumption that we are dealing with an electron in a spherically symmetric initial state, $\psi_0=\psi_0(|\v x|)$, with the energy $E_0$ which after excitation becomes a plane wave $\psi_{\v q}=\exp(i\v q \v x)$, and thus $d\mu(\v q)=d^3q$ and a fixed value of \mbox{$\wq=q^2/2m^\star-E_0$} translates into a fixed value of $q=|\v q|$. In a real semiconductor both of the functions $\psi_0$, $\psi(\v q)$ are different but, of course, for a given crystal they are very well known.
With these assumptions we find
\begin{equation}
d^i(\v q)=e q^i h(q),
\end{equation}
with a certain function $h(q)$. The $d^3q$ integration is now straightforward; with the integral over the directions, $\int d\Omega\, q^iq^j=\tfrac{4\pi}{3}\delta^{ij}\, q^2$, we finally get for the leading contribution from $\int d^2g\, F\star E$:
\begin{equation}\label{lead}
\int d^2g\, F(\tau)\star E(s)=\frac{16e^2\pi^3}{3}  m^\star q_0^3|h(q_0)|^2\, K^i\, \int d^3p\, \delta(p-\w) b^*_i(\v p) e^{ip\th}
\end{equation}
where $q_0=\sqrt{2m^\star (\w+E_0)}$. This expression is equivalent to
\begin{equation*}
\int d^2g\, F(\tau)\star E(s)=A(\w)\, K^i\, E^{-}_i(\th)|_{\w}
\end{equation*}
with $E^-_i(\th)|_{\w}$ meaning the negative frequency part of the electric field operator restricted to the frequency $\w$, taken at the time $\th$, and with the abbreviation $A(\w)=\frac{16e^2\pi^3}{3}  m^\star q_0^3|h(q_0)|^2$. Analogously we find
\begin{equation*}
\int d^2g\, E(\tau)\star F(s)=A(\w)\, K^i\, E^{+}_i(\th)|_{\w}.
\end{equation*}
The expectation value of the output of the BHD detector, $\langle
J\rangle_{(S,F)}$, eq. \eqref{av_J}, is therefore given  to the leading order by
\begin{equation}
    \langle J\rangle_{(S,F)}=A(\w) K^i\ \big\langle  E_i(\th,\ax)|_\w+ E_i(\th,\ay)|_\w\big\rangle_S.
\end{equation}

\subsubsection{Limit $T\ra \infty$ and monochromatic LO}
It must be stressed, that the limit $T\ra \infty$ taken here with the assumption of monochromaticity of the local oscillator field, which leads to \eqref{lead}, is not physical and mathematically troublesome (intergating the response for infinitely long times / smearing with singular test functions leading to products of delta functions for products of operators of the type \eqref{lead}). To avoid these troubles one should replace $K^i$ by a frequency-dependent function $k_i(\w)$ and integrate \eqref{lead} over $\w$. The function $k_i(\w)$ can be for instance such as to correspond to ``pulsed'' almost-monochromatic local oscillators; the equation \eqref{lead} is then replaced by
\begin{equation}\label{smeared}
\int d^2g\, F(\tau)\star E(s)=A(\w)\, \int d^3p\, k^i(p) b^*_i(\v p) e^{ip\th}
\end{equation}
with a smooth $k^i(p)$ sharply peaked around the frequency $\w$ of the local oscillator.

\subsubsection{Generalization to non-trivial geometries or inhomogeneous media}
For quantum fields in non-trivial static environments (waveguides, cavities or materials with position-dependent $\epsilon(\v x)$ and $\mu(\v x)$), the electric-field operators possess a generalization of the decomposition \eqref{electric_field}
\begin{equation}
E_i(t,\ax)=\int d\nu(p^a)\, \left[ e^{-i\w(p^a)t}
 b(p^a)\psi_i(p^a, \ax)+e^{i\w(p^a)t}b^*(p^a)\ba{\psi_i(p^a,\ax)}\right],
\end{equation}
where the multi-index $p^a$ corresponds to the parameters of electromagnetic waves supported by the environment (such as types of waves and their wave vectors), $\psi_i(p^a, \ax)$ denotes the (electric field of the) solution of the Maxwell equations with these parameters and $d\nu(p^a)$ denotes the appropriate measure. The frequency, $\w(p^a)$, is given by an appropriate dispersion relation. The functions $\psi_i(p^a, \ax)$ are normalized so that $[b(p^a),b^*(k^a)]=\de(p^a-k^a)$. The generalization of eq. \eqref{smeared} is then
\begin{equation}
\int d^2g\, F(\tau)\star E(s)=A(\w)\, \int d\nu(p^a) \, k^i[\w(p^a)] b^*(p^a) \psi_i(p^a,\ax)\equiv A(\w) E^{-}(\th)|_{k(\w)},
\end{equation}
which is  the negative-frequency part of the field operator restricted in frequencies and polarizations by $k^i(\w)$. This seemingly straightforward generalization has the profound effect of making the commutator of positive and negative frequency part of the field operators position ($\ax)$ dependent. As a consequence there may be reduced fluctuations of \emph{both} so-called quadratures of the quantum field. An example of this is presented in the appendix \ref{Casimir}.

\subsubsection{Variance of the BHDs output}
For many interesting states
$S$, the expectation value of the electric field operator (and thus the leading term of $\langle J\rangle_{(S,F)}$) vanishes, for instance
if $S$ is the ground/vacuum state, or an eigenstate of the photon-number operator, or a superposition of pairs of photons - eg. the squeezed
state. Let us present the expectation value of the square of $J$ in the ideal case of monochromatic LO and homogeneous media. The calculation is trivial as we manipulate the operators, under the expectation value, which are a product of two independent expressions. The leading term of  $\langle J^2\rangle $ is
given by
\begin{equation}\label{J_squared}
    \langle J^2\rangle_{(S,F)}=A^2(\w)\, K^i\,K^j \Big\langle  \bigl[E_i(\th,\ax)|_\w+ E_i(\th,\ay)|_\w\bigr]\bigl[E_j(\th,\ax)|_\w+ E_j(\th,\ay)|_\w\bigr]\Big\rangle_S.
\end{equation}

\section{Discussion}\label{remarks}

\subsection{Properties of $\langle J^2\rangle$}
We note the following:
\begin{itemize}
\item $\langle J^2\rangle$ is proportional to $K^2$, that is to the power of the local oscillator field; by taking measurements with growing values of $K^2$  the value of the fluctuations of the electric field in the state $S$ can be quantitatively estimated. This is the standard signature checked for in experiments, see eg. \cite{pulsed,mr}.

\item Should one wish to compare $\langle J^2\rangle$ of two different states on the decibel scale, the dependence on the yet unspecified parameter, $A$, and LO power drops out, and the result depends only on the (frequency-restricted) two-point functions of the states under consideration.

\item If the state $S$ can be assumed to have certain symmetries, a further simplification of the form of $\langle J^2\rangle$ is possible. For instance if $S=\Omega$ is the Poincare invariant (usual) vacuum, then
\begin{equation}
    \langle J^2\rangle_{(S,F)}=2 A^2(\w)\, K^i\,K^j \Big\langle  E_i(\th,\ax)|_\w\, E_j(\th,\ax)|_\w\Big\rangle_S.
\end{equation}
(The spatial orientation of the photodiodes plays a role here.)
\item All the operators are restricted to the frequency, $\w$, of the LO; therefore time is here $2\pi/\w$ periodic, and thus $\th\in[0,2\pi/\w)$ is the measure of time with $2\pi/\w$ corresponding to the period of the LO wave.
\end{itemize}

\subsection{Interaction Hamiltonians}\label{rem2}

With regard to the interaction Hamiltonian which has been employed, \eqref{f_dipol}, we note that we could have taken it without the restriction on the electric-field operator, eq. \eqref{hamiltonian}; its frequencies would nonetheless get restricted to the frequency of LO (at the BHD) and the large wavelength of light waves of this frequency (against the essential support of $\psi_0$) would lead to the same formulas for the output of the BHD. Moreover, because of the standard relation
$$\bigl(\psi(\v q), \, p^j\, \psi_0 \bigr)=i m^\star \w_{\v q} \cdot \bigl(\psi(\v q), \, x^j\, \psi_0 \bigr),$$
 which is generally valid if $\psi_0$ and $\psi(\v q)$ are eigenstates of the Hamiltonian $H_0$,  the same expressions for the expectation values of $J$ and $J^2$ would have been obtained had we started with the interaction Hamiltonian of the form $\tfrac{e}{m^\star}A^i(t,\v x)\, p_i$.

\subsection{Note on the locality of measurements}
Let us assume that $\psi_0(\v x)$ is supported in a compact region\footnote{We ignore the typical exponential tails of the wavefunctions.} and that $g(t)$ is also of a compact support (in time). Then the quantum-field-theoretical observable corresponding to $\text{PD}(g,\ax)$ is just an integral (over $d\mu(\v q)$) of a product of two electric field operators smeared with test functions supported in the spacetime region $\text{supp }\psi_0 \times\text{supp } g$. Such an observable is QFT-localized in the smallest double-cone\footnote{Double cones are causally-complete regions of the form $J^+(x)\cap J^-(y)$.} containing this region. As a consequence, two such observables localized in causally separated double-cones will commute (and thus have their probabilistic distributions unconstrained by Heisenberg-like uncertainty relations). Moreover, since the operator corresponding to $\text{PD}(g,\ax)$ is clearly positive, it cannot annihilate the vacuum because otherwise  by the theorem of Reeh and Schlieder\cite{CH,Haag} it would necessarily be a null operator (vacuum is separating for the local observables). Thus, strictly, photodiodes are not ``photon counters'', as they exhibit non-null vacuum response. These ``vacuum-effects'' were also encountered in \cite{FGO,DO}, and are related to causality problems of Hagerfeldt \cite{Ha} (see also the subsequent clarification by Buchholz and Yngvason \cite{BY}).

\section{Conclusions and outlook}

We have corroborated the result, that the output of a balanced homodyne detector provides a quantitative measure of the n-point functions of the states of the quantum electric field.
Moreover we have shown how this result can be generalized to the case of detectors placed in inhomogeneous environments. Non-trivial effect are expected in such a case, as is shown in the appendix A dealing with quantum fields and BHDs outputs in Casimir cavities (see also \cite{BM,HJSV,F,SF}). Our investigation supports the view, that BHD-like detectors are suitable for measurements of intriguing QFT effect such as negative energy densities present for instance for squeezed states, and thus provide a tool for testing recent theoretical predictions associated with Quantum Energy Inequalities \cite{FR,appl}.

\begin{acknowledgments}
I would like to thank K. Fredenhagen for encouraging discussions in the initial phase of this work. I am also indebted to T. Roman and L. Ford for
discussions and for drawing my attention to the references \cite{F,HJSV}. The financial support of the Erwin Schr\"odinger Institute, where this work was finalized, is also greatfully acknowledged.
\end{acknowledgments}

\appendix
\section{Balanced homodyne detectors in Casimir setups}
\label{Casimir}
In this appendix we will derive a prediction for the output of a BHD placed between Casimir plates. In order to do so we will recall the results of \cite{HJSV} and provide their relation to the characteristics of the output of the detector. In ref. \cite{HJSV} autocorrelation-functions (two-point-functions) were derived for the ground and thermal state of the quantum electromagnetic field between perfectly conducting parallel plates. Let us take the ground state $G$ for simplicity. The autocorrelation function for the electric field operators, $\langle E_i(x)E_j(x')\rangle_G$ is obtained (Eq. (2.25c) of \cite{HJSV}) by partially differentiating (w.r.t. the first argument) the image-sum-functions
\[
F^{\mp}(x,x')=-\frac{1}{4\pi^2}\sum_{n=-\infty}^\infty \frac{1}{(x\mp x'-n L)^2+(y-y')^2+(z-z')^2-(t-t')^2}
\]
where $L$ is \emph{twice} the distance between the plates which are perpendicular to the $x$-direction. Specifically for electric field operators in the $y$-direction (tangential to the plates) we have
\[
\langle E_y(x)E_y(x')\rangle_G=(\d_x^2+\d_z^2)\left[F^-(x,x')-F^+(x,x')\right],
\]
where differentiations are performed with respect to the first point ($x$) only.
Let us now set the points so, that $x$ lie at the same spatial position as $x'$ but by $s$ later in time: $x=(s,\X,0,0)$, $x'=(0,\X,0,0)$, where $\X$ is the distance from the first pate.  We find
\[
 \langle E_y(x)E_y(x')\rangle_G=\frac{1}{\pi^2}\sum_{n=-\infty}^{\infty}
 \left[
 \frac{(nL)^2+s^2}{(s-nL)^3(s+nL)^3}-\frac{(nL-2\X)^2+s^2}{(s-nL+2\X)^3(s+nL-2\X)^3}
 \right]
\]
which is a function of $s$ and $\X$. The Fourier transform of this  function with respect to $s$ gives the frequency spectrum (eq. (2.30) of ref. \cite{HJSV}),
\[
 \s_G(\w,\X)=\frac{1}{2\pi}\int_{-\infty}^\infty ds\, e^{i\w s}\,  \langle E_y(x)E_y(x')\rangle_G= \frac{\w^3}{4\pi^2} \sum_{n=-\infty}^{\infty}\bigl\{Q[\w n L]-Q[\w(nL-2\X)]\bigr\},
\]
where the function $Q$ is given by
\[
Q(x)=\frac{\sin x}{x}+\frac{\cos x}{x^2}-\frac{\sin x}{x^3}.
\]
We note, that the $n=0$ term of the first summand, equal to $\w^3/6\pi^2$ corresponds to the Minkowski-vacuum, $\Omega$, spectral density (i.e. to the situation without plates). Hacyan et al. have summed up the series and expressed the $n$-sums in terms of piecewise elementary functions (polynomials and trigonometric functions of $\w$). For the purpose of this paper let us merely plot vacuum and Casimir spectral densities for $L=2\mu m$ (plate separation $1\mu m$).
\begin{figure}[ht]\centering
  \includegraphics[scale=0.55]{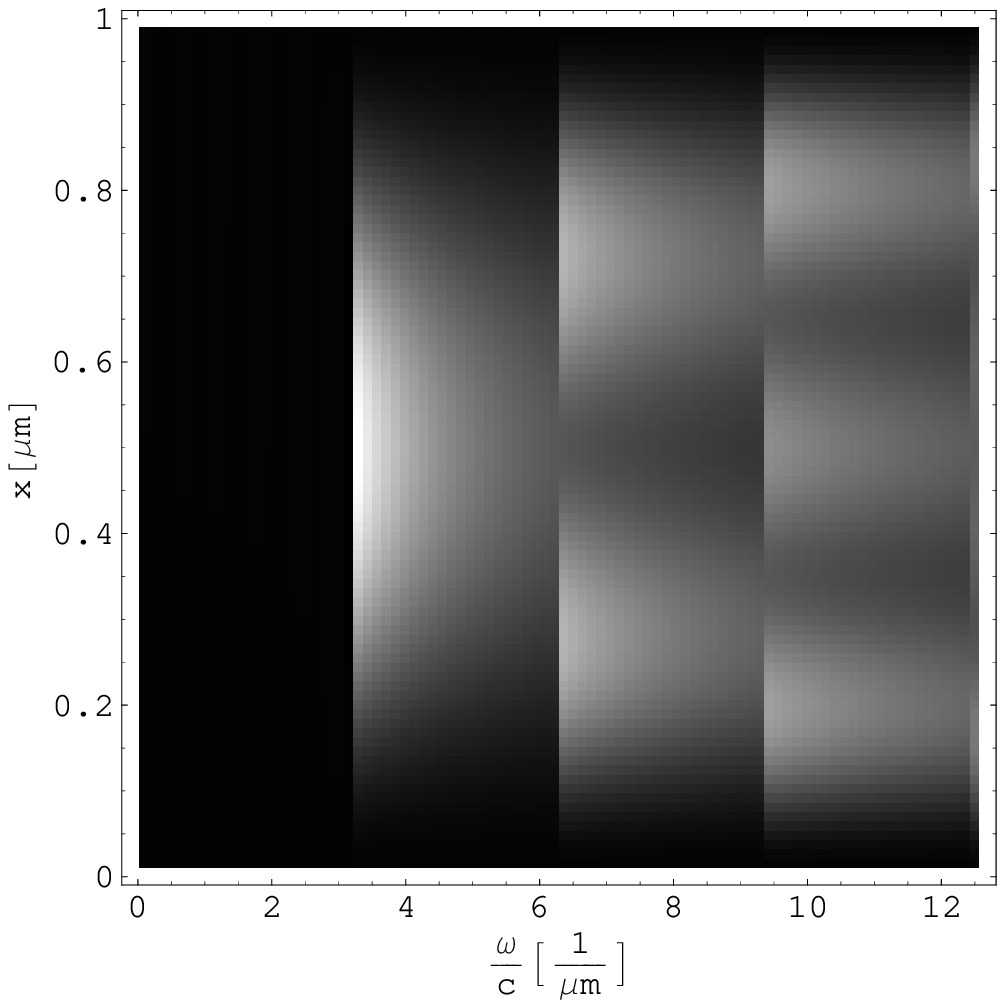}
\caption{The normalized difference $({\s_G(\w,\X)}-{\s_\Omega(\w,\X)})/\s_\Omega(\w,\X)$ between vacuum and Casimir spectral densities for $\w/c\in[0,4\pi]\mu m^{-1}$. Black color corresponds to negative values (sub-vacuum spectral densities). For $\w L<2\pi$ Casimir spectral density vanishes, $\s_G(\w,\X)=0$, for all $\X$. Discontinuities appear at $\w L=2n\pi$.}
\end{figure}

\begin{figure}[ht]\centering
  \includegraphics[scale=0.75]{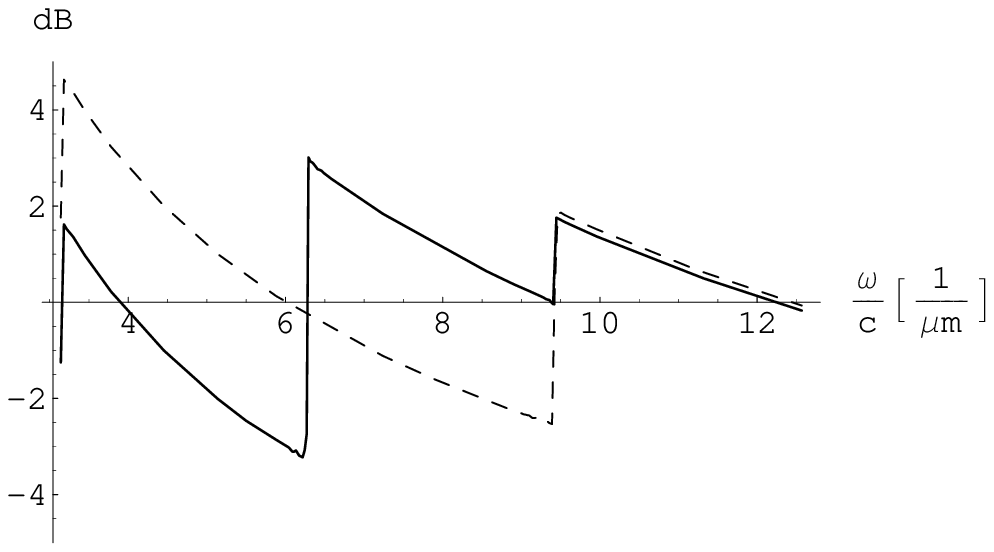}
\caption{Frequency-dependence of the suppression of fluctuations in the dB scale, \mbox{$10\cdot \log_{10}\left[{\s_G(\w,\X)}/{\s_\Omega(\w,\X)}\right]$}, for the ground state in the Casimir geometry. Plotted curves correspond to $x=0.25\mu m$ (solid) and $x=0.5\mu m$ (dashed).}
\end{figure}

Let us finally make a connection to the observables measured by BHD's. For quantum fields in non-trivial static environments we find, that (in the notation of section II.B.3)
\begin{equation}
\s_G(\w,\X)=\int d\nu(p^a)\, d\nu(k^a)\, \delta\left[\w(p^a)-\w\right] |\psi_y(p^a,\X)|^2 \delta(p^a-k^a).
\end{equation}
The expectation value of the output of the BHD in the ground state vanishes, while its variance, \eqref{J_squared}, contains two parts, one of which depends on the relative position of the diodes (this part comes from the term $\langle E_i(t,\ax)E_j(t,\ay)\rangle$). The position-\emph{independet} part is simply related to the spectral density:
\begin{align}
\langle J^2 \rangle_{indep}&=A^2(\w_{LO}) \Big\langle  E(\th,\ax)|_{k(\w)}E(\th,\ax)|_{k(\w)}+ E(\th,\ay)|_{k(\w)}E(\th,\ay)|_{k(\w)}\Big\rangle_S
=\\&=2 A^2(\w_{LO})\int d\w\, \s(\w,\X) |k^y(\w)|^2,
\end{align}
where $\w_{LO}$ is the local-oscillator frequency. Thus in the usual situation where $k^y(\w)$ is sharply peaked around $\w_{LO}$ the spectral densities presented above are proportional to the position-independent (and LO-phase, $t_0$, independent) output of a balanced homodyne detector.

\section{Properties of the smearing functions}
In the paper, the arbitrary functions of time, $g(t)$, multiplied the interaction Hamiltonian. These functions should be set to one in the spacetime region, where the actual interaction takes place. Thus we view them as being smooth and essentially equal to one in the interval $[-T/2,T/2]$, where $T$ is taken sufficiently large. Let us take for instance $$g(t)=\frac{1}{2}\left\{\tanh[a(t+T/2)]-\tanh[a(t-T/2)]\right\}.$$
we find easily (using the convention $\hat g(\w)=\int e^{i\w t} g(t)\, dt$),
\begin{equation*}
\widehat{\tanh} (\w)=\frac{i\pi}{\sinh(\pi\w/2)},\\
\end{equation*}
and
\begin{equation*}
\hat g (\w)=\frac{\pi}{a}\, \frac{\sin(\w T/2)}{\sinh(\pi\w/2 a)}.
\end{equation*}
This function evidently decays rapidly for $\w$ large with respect to $a$. Moreover $\hat g(0)=T$, and due to the known representation of the Dirac's delta
\begin{equation*}
\lim_{T\ra \infty}\frac{\sin(\w T)}{\w}=\pi\delta(\w)
\end{equation*}
we have
\begin{equation*}
\lim_{T\ra\infty} \hat g(\w)=2\pi \delta(\w).
\end{equation*}
This property is generic; any positive real function of time, $g_T(t)$, which converges weakly (on the space of functions of rapid decay) to one as the parameter $T\ra \infty$ has the property  $\hat g_T(\w)\ra 2\pi\de(\w)$. If $g_T(t)$ is smooth, $\hat g_T(\w)$ will be of rapid decay (for finite $T$).

\end{document}